\documentclass[apj]{emulateapj}
\usepackage{multirow}
\usepackage{graphics}
\usepackage{rotating}
\usepackage{enumitem}
\usepackage{amsmath}
\usepackage{txfonts}
\usepackage{hyperref}
\usepackage{setspace}
\usepackage{float}
\usepackage{tablefootnote}
\usepackage{footnote}
\usepackage{threeparttable}
\usepackage[mediumspace,mediumqspace,Grey,squaren]{SIunits}

\hypersetup{
			hyperfootnotes=true,
			colorlinks=true,
			linkcolor=blue,
			citecolor=blue,
			urlcolor=blue,
			breaklinks=false,
}

\slugcomment{}
\shorttitle{Compact UV Size of Green Peas}
\shortauthors{Kim et al.}

\begin{document}

\title{The Compact UV Size of Green Pea Galaxies As Local Analogs of High-redshift Ly$\alpha$-Emitters}
\author{Keunho J. Kim\altaffilmark{1,2}, Sangeeta Malhotra\altaffilmark{2,3}, James E. Rhoads\altaffilmark{2,3}, and Huan Yang\altaffilmark{4}}

\altaffiltext{1}{Department of Physics, University of Cincinnati, Cincinnati, OH 45221, USA; kim2k8@ucmail.uc.edu}
\altaffiltext{2}{School of Earth $\&$ Space Exploration, Arizona State University, Tempe, AZ 85287, USA}
\altaffiltext{3}{NASA Goddard Space Flight Center, Greenbelt, MD 20771, USA}
\altaffiltext{4}{Las Campanas Observatory, Carnegie Institution for Science, Chile}

\def\OI{[\mbox{O\,{\sc i}}]~$\lambda 6300$}
\def\OIII{[\mbox{O\,{\sc iii}}]~$\lambda 5007$}
\def\OIIIs{[\mbox{O\,{\sc iii}}]~$\lambda 4363$}
\def\OIIIab{[\mbox{O\,{\sc iii}}]$\lambda\lambda 4959,5007$}
\def\SIIab{[\mbox{S\,{\sc ii}}]~$\lambda\lambda 6717,6731$}
\def\SII{[\mbox{S\,{\sc ii}}]~$\lambda \lambda 6717,6731$}
\def\NII{[\mbox{N\,{\sc ii}}]~$\lambda 6584$}
\def\NIIb{[\mbox{N\,{\sc ii}}]~$\lambda 6584$}
\def\NIIa{[\mbox{N\,{\sc ii}}]~$\lambda 6548$}
\def\NI{[\mbox{N\,{\sc i}}]~$\lambda \lambda 5198,5200$}

\def\OIIa{[\mbox{O{\sc ii}}]~$\lambda 3726$}
\def\OIIb{[\mbox{O{\sc ii}}]~$\lambda 3729$}
\def\NeIIIa{[\mbox{Ne{\sc iii}}]~$\lambda 3869$}
\def\NeIIIb{[\mbox{Ne{\sc iii}}]~$\lambda 3967$}
\def\OIIIa{[\mbox{O{\sc iii}}]~$\lambda 4959$}
\def\OIIIb{[\mbox{O{\sc iii}}]~$\lambda 5007$}
\def\HeII{{He{\sc ii}}~$\lambda 4686$}
\def\ArIVa{[\mbox{Ar{\sc iv}}]~$\lambda 4711$}
\def\ArIVb{[\mbox{Ar{\sc iv}}]~$\lambda 4740$}
\def\NIa{[\mbox{N{\sc i}}]~$\lambda 5198$}
\def\NIb{[\mbox{N{\sc i}}]~$\lambda 5200$}
\def\HeI{{He{\sc i}}~$\lambda 5876$}
\def\OI{[\mbox{O{\sc i}}]~$\lambda 6300$}
\def\OIb{[\mbox{O{\sc i}}]~$\lambda 6364$}
\def\SIIa{[\mbox{S{\sc ii}}]~$\lambda 6716$}
\def\SIIb{[\mbox{S{\sc ii}}]~$\lambda 6731$}
\def\ArIII{[\mbox{Ar{\sc iii}}]~$\lambda 7136$}

\def\Ha{{H$\alpha\,$}}
\def\Hb{{H$\beta\,$}}

\def\NIIHa{[\mbox{N\,{\sc ii}}]/H$\alpha$}
\def\SIIHa{[\mbox{S\,{\sc ii}}]/H$\alpha$}
\def\OIHa{[\mbox{O\,{\sc i}}]/H$\alpha$}
\def\OIIIHb{[\mbox{O\,{\sc iii}}]/H$\beta$}

\def\Ebmv{E($B-V$)}
\def\LOIII{$L[\mbox{O\,{\sc iii}}]$}
\def\Ledd{${L/L_{\rm Edd}}$}
\def\LOIIIs4{$L[\mbox{O\,{\sc iii}}]$/$\sigma^4$}
\def\LOIIIMbh{$L[\mbox{O\,{\sc iii}}]$/$M_{\rm BH}$}
\def\Mbh{$M_{\rm BH}$}
\def\Msigma{$M_{\rm BH} - \sigma$}
\def\Ms{$M_{\rm *}$}
\def\Msun{$M_{\odot}$}
\def\Msunyr{$M_{\odot}yr^{-1}$}
\def\bt{$B/T\,$}
\def\btr{$B/T_{\rm r}\,$}

\def\ergs{$~\rm ergs^{-1}$}
\def\kms{${\rm km}~{\rm s}^{-1}$}
\newcommand{\cms}{\mbox{${\rm cm\;s^{-1}}$}}
\newcommand{\pccm}{\mbox{${\rm cm^{-3}}$}}
\newcommand{\sauron}{{\texttt {SAURON}}}
\newcommand{\oasis}{{\texttt {OASIS}}}
\newcommand{\HST}{{\it HST\/}}

\newcommand{\Vg}{$V_{\rm gas}$}
\newcommand{\Sg}{$\sigma_{\rm gas}$}
\newcommand{\eg}{e.g.,}
\newcommand{\ie}{i.e.,}

\newcommand{\gandalf}{{\texttt {gandalf}}}
\newcommand{\fracDeV}{{\texttt {FracDeV}}} 
\newcommand{\ppxf}{{\texttt {pPXF}}}

\newcommand{\sersic}{S\'{e}rsic}
\newcommand{\lumpc}{$L_{\rm bol, 250pc}$}
\newcommand{\lumtot}{$L_{\rm bol, total}$}
\newcommand{\sbpc}{$S_{\rm 250pc}$}
\newcommand{\sbeff}{$S_{\rm eff}$}
\newcommand{\reff}{$R_{\rm eff}$}
\newcommand{\rtotal}{$R_{\rm total}$}
\newcommand{\ewlya}{EW(Ly$\alpha$)}
\newcommand{\fesclya}{$f^{Ly\alpha}_{esc}$}
\newcommand{\lya}{Ly$\alpha$}
\newcommand{\prad}{$r_{\rm cir,50}$}
\newcommand{\lyalum}{$L(\rm{Ly}\alpha)$}

\begin{abstract}
We study the dependence of \lya\ escape from galaxies on UV continuum size and luminosity using a sample of 40 Green Pea (GP) galaxies, which are the best local analogs of high-redshift \lya\ emitters (LAEs).
We use the Cosmic Origins Spectrograph near-ultraviolet images from the \textit{Hubble Space Telescope} to measure the UV size and luminosity with $0.047''$ spatial resolution.
Like most galaxies the GPs show a log-normal size distribution.
They also show a positive correlation between size and UV-continuum luminosity.
The slope of the size-continuum luminosity relation for GPs is consistent with those of continuum-selected star-forming galaxies at low and high redshifts. A distinctive feature of GPs is a very compact typical radius of 0.33 kpc with a population spread (1$\sigma$) of 0.19 kpc.
The peak of the size distribution and the intercept of the size-luminosity
relation of GPs are noticeably smaller than those of continuum-selected star-forming
galaxies at similar redshifts. There are statistically significant anticorrelations found between the circularized half-light radius (\prad), the \lya\ equivalent width (\ewlya), and the \lya\ escape faction (\fesclya), suggesting that small UV-continuum radii are crucial for \lya\ emission.  GPs and high-redshift LAEs have similar sizes, once spatial resolution effects are properly considered. Our results show that a compact small size is crucial for escape of \lya\ photons, and that \lya\ emitters show constant characteristic size independent of their redshift.
\end{abstract}	


\section{Introduction}
\label{sec:introduction}
\lya -emitting galaxies (i.e., \lya -emitters, LAEs) are a class of galaxies characterized by spectra with prominent \lya\ emission lines.  Since \lya\ photons are usually generated by intense star formation activity, most LAEs have relatively high star formation rate (SFR) for their stellar mass (i.e., specific star formation rate (sSFR) $\gtrsim 10^{-8} \rm{yr^{-1}}$) and relatively young stellar population ages ($\lesssim 50 \ \rm{Myr}$) \citep[e.g.,][]{malh02,gawi07,pirz07,fink15a,sant20}. However, because \lya\ photons are resonantly scattered by neutral hydrogen and significantly affected by dust absorption/scattering \citep[e.g.,][]{ahn03}, not all actively star-forming galaxies (SFGs) have observable \lya\ emission, which requires that \lya\ photons manage to escape from star-forming regions within the galaxy all the way to the intergalactic medium (IGM).  So, what are the properties of LAEs that allow \lya\ to escape?

\lya\ emission from galaxies is also one of the most promising indicators for identifying Lyman-continuum (LyC) leakers (i.e., galaxies that emit significant
amounts of ionizing radiation to the IGM) \citep{verh15,deba16,izot16,gaza20}.  LyC leaking galaxies are important for cosmology, because they were significant and probably dominant sources of the photons that drove reionization of the early universe
($z > 6)$.   Thus, studying the physical properties of LAEs and understanding their \lya\ (and potentially LyC) escape mechanisms have been important topics in the fields of galaxy evolution and cosmology, respectively \citep[e.g.,][and references therein]{malh02,malh12,rhoa00,rhoa14,fink15a,oyar17,rive17,runn20}.

Multiple observations and morphological analysis of LAEs show that they are mostly compact (with effective radius $r_{\rm eff} \lesssim 1.5$ kpc), sometimes with clumpy features shown in UV continuum, at high redshifts $2 \lesssim z \lesssim 7$ \citep[e.g.,][]{dowh07,over08,bond09,tani09,bond12,malh12,jian13,paul18,shib19,rito19}. 
Based on the approximately constant typical sizes of LAEs over a wide span of redshift ($2 \lesssim z \lesssim 6$), \citet{malh12} suggested that the compact size of LAEs is a crucial physical condition for a galaxy to become an LAE, as recent analytic calculations on morphologically compact conditions for LyC leakers also suggest \citep{cen20}. This has been supported by later studies \citep[e.g.,][] {jian13,paul18}, but others \citep[notably][] {shib19} have challenged this with the view that LAEs follow similar size-luminosity relations, and show size evolution with redshift similar to high-redshift Lyman Break galaxies.
Among the Lyman Break samples, moreover, it has been suggested that size evolution differs between brighter ($L_{\rm UV}>0.3 L_*$) and fainter ($L_{\rm UV}<0.3 L_*$) galaxies, where $L_{\rm UV}$ and $L_{*}$ are the UV luminosity and the characteristic luminosity of galaxies, respectively \citep[e.g.,][]{bouw15,fink15b}).

Extending this redshift dependence to lower redshifts would give a more sensitive  test of whether the characteristic radius of LAEs evolves like $(1+z)^{-1.37}$ \citep{shib19} or is flat. Going from $z=6$ to 2 gives a factor of 3 change in the size while going to $ z=0.3$ predicts more than a factor of 10 in radius change.  Green Peas are the best low-redshift analogs of high-redshift LAEs in terms of \lya\ equivalent width distribution \citep{yang16}.   Green Peas were selected as high equivalent width [OIII] emitters in a citizen science project, and have generally low metallicities and low masses  \citep{card09,amor10,jask13,henr15,izot16,yang16,yang17a,yang17b,izot18a,orli18,jask19, jian19a,jian19b,hoga20,brun20,clar21}.

This is the second in a series of papers where  we investigate the UV continuum morphologies of Green Pea (GP) galaxies  to understand the mechanisms enabling \lya\ escape. In the first paper \citep[][hereafter Paper I]{kim20} we investigated the effects of star-formation per unit area (star formation intensity, SFI) and specific star-formation rate (sSFR).  In this paper we investigate the UV size and luminosity of a sample of GPs that were observed with the Cosmic Origins Spectrograph near-ultraviolet (COS/NUV) acquisition images from the \textit{Hubble Space Telescope (HST)}. The high angular resolution of the COS/NUV images (0.0235 arcsec $\rm{pixel}^{-1}$ sampling and PSF FWHM of 0.047 arcsec) and the low redshifts ($z =$ 0.1-0.3) of GPs have enabled us to measure the spatially resolved UV sizes with a factor of $\sim 2$ better resolution limit compared to other \textit{HST} instrument+filter combinations typically used for measuring the sizes of high-\textit{z} LAEs  (i.e., 0.047 arcsec, vs. 0.09 arcsec for the Advanced Camera for Surveys (ACS) F850LP filter, and 0.15 arcsec for the Wide Field Camera 3 (WFC3) F160W) \citep{wind11}.

We use this exquisite spatial resolution to test whether low-redshift LAEs have compact physical sizes, which would be expected if the small sizes of high-$z$ LAEs are causally linked to \lya\ escape.   Alternatively, if high-$z$ LAEs are compact merely because high-$z$ galaxies are compact in general, we would expect their local counterparts to be larger.
We also use our data set to study more general aspects of the relation between \lya\ escape and UV continuum size.

Section \ref{sec:samples and data analysis} describes our galaxy sample and procedures for UV size and luminosity measurements. In Section \ref{sec:results}, we present our results.  We summarize our conclusions in Section \ref{sec:Summary and Conclusions}.  Throughout this paper, we adopt the AB magnitude system and the $\Lambda$CDM cosmology of ($H_{0}$, $\Omega_{m}$, $\Omega_{\Lambda}$) = (70 $\rm{kms^{-1}}$ $\rm{Mpc^{-1}}$, 0.3, 0.7).

\section{SAMPLE AND DATA ANALYSIS}
\label{sec:samples and data analysis}

\subsection{GP sample}
\label{subsec:Sample selection}
Our GP sample consists of the 40 galaxies with UV spectra that are analyzed and presented by Yang et al. 2017b (hereafter Y17) and Paper I. Our sample GPs have redshifts $0.1 \lesssim z \lesssim 0.35$. Green Peas  were selected by citizen scientists for colors indicating a strong [O III] $\lambda5007$ emission line in the SDSS \textit{r}-band \citep{card09} and compact unresolved morphology at SDSS seeing of 1.2-2". The current sample excludes galaxies whose line ratios indicate an active galactic nucleus, based on the BPT diagram (Baldwin et al. 1980). As we will see in the next section, the measured NUV sizes of the GPs are much smaller than the SDSS seeing. Therefore, the citizen science criteria that GPs be ``round'' and ``pea-like'' do not bias the measured median sizes much.

The \textit{HST} UV spectroscopy was drawn from several observing programs, but notably, about half this sample was new in Y17, and was selected to sample the range of metallicity and dust extinction as much as possible.
Of the 43 GPs in Y17, we removed 3 galaxies--- one (GP ID 0747+2336) that shows no \lya\ emission line, and two others (GP ID 0021+0052 and 0938+5428) where the COS imaging data were obtained using the MIRRORB instrument  configuration\footnote{$\rm{http://www.stsci.edu/hst/cos/documents/isrs/ISR2010}$\_10.pdf}. We refer to Y17 for further details on sample selection procedures.

We adopt GP ID and \lya\ properties such as the equivalent width of \lya\ line (\ewlya), the \lya\ escape fraction (\fesclya), and the \lya\ luminosity (\lyalum) from Y17. We also adopt galactic internal extinction ($A_{\rm int}$), Milky Way extinction ($A_{\rm MW}$), and the $k$-correction ($k$) values from Paper I to derive the extinction and $k$-corrected UV luminosity at rest-frame wavelength of 1877 $\angstrom$ of our sample GPs in Section \ref{subsec:Size and Luminosity measurements}. 

As described in Paper I, the 40 sample GPs have been observed with the COS/NUV acquisition images with the MIRRORA configuration. The pivot wavelength of the observed filter is 2319.7 $\angstrom$. The exposure time of the NUV images of our sample GPs is mostly greater than 100 s. For more details we refer the reader to Paper I.

\subsection{Size and Luminosity Measurements}
\label{subsec:Size and Luminosity measurements}
We measure the UV-continuum size and luminosity of our sample GPs using two different size measures: first, the Petrosian half-light radius \citep{petr76}; and second, the effective radius measured from surface brightness profile fitting using the public software \textsf{GALFIT} (version 3.0.5) \citep{peng02,peng10}. 
The Petrosian radius has certain advantages (described below) that led us to use it as our size measure of choice for most analyses in this study.   The exception is Section 3.4, where we compare the typical size of GPs with those of high-redshift LAEs by employing the effective radius to ensure consistency between low- and high-redshift size measures\footnote{We note that there is a systematic size difference between the two size measures (i.e., Petrosian half-light radius and the effective radius) depending on the concentration of the galaxy's light profile (i.e., S\'{e}rsic index) \citep[][]{grah05a,grah05b}.  We checked that both size measures of our sample GPs give qualitatively similar results throughout our analysis, though the effective radius has a broader size distribution with a higher median size compared to the Petrosian half-light radius (i.e., the median effective radius is $\sim$ 0.62 kpc, while the median Petrosian half-light radius is $\sim$ 0.33 kpc).}.

To measure the sizes, we first derive the galaxy center and apparent axis ratio $b/a$ (where $a$ and $b$ indicate the semi-major and semi-minor axis of a galaxy, respectively, and ellipticity $\epsilon = 1 - b/a$) by performing GALFIT surface brightness profile fitting. We adopt the same sky-subtracted 140 $\times$ 140 pixel (3.3$''$ $\times$ 3.3$''$) COS/NUV images of our sample GPs and the point-spread function (PSF) image of star P330E observed during the \textit{HST} program 11473\footnote{$\rm{http://www.stsci.edu/hst/cos/documents/isrs/ISR2010}$\_10.pdf} that we processed in Paper I to measure the central star star formation intensity (SFI, star formation rate per unit area) of this sample.  During the fitting procedure, we assume one component S\'{e}rsic profile, and the associated background noise $\sigma$ images are internally generated by considering image information such as exposure time, readout noise, and the number of combined images from the image header. The initial guesses for galaxy center, apparent magnitude, and the effective radius are adopted from the measured values in Paper I. Specifically, the brightest pixel of the NUV images, and the Petrosian-based galaxy segmentation maps and the pixel-count based size parameter \citep{petr76,law07,ribe16} are used to obtain initial values of the galaxy center, apparent magnitude, and effective radius. The initial values for $b/a$ and Position Angle (i.e., the direction of galaxy major axis in the sky) are set 0.8 and 0 degree, respectively, for all sample galaxies.

Additionally, we put constraints on the fitting range of some parameters. A fitting solution for this type of multiparameter fit does not always guarantee a physically sound result, even with a reasonably good reduced-$\chi^{2}$ value \citep[e.g.,][for relevant discussions]{peng10,meer15,kim16}. Therefore, to make sure that the fits are physically motivated, the fitted galaxy center is to be within 5 pixels from the initial galaxy center; the effective radius is greater than 1 pixel but less than the twice the total radius derived from the galaxy size parameter mentioned above; the S\'{e}rsic index is between 1 and 8; and the $b/a$ value is fitted between 0.2 to 1.

From the fit results, we employ the galaxy center, apparent axis ratio ($b/a$), and the Position Angle to measure the Petrosian radius and luminosity. Since the Petrosian radius depends on the surface brightness profile (i.e., a curve of growth) of a galaxy, this size definition has the advantages of being redshift independent, less sensitive to dust reddening, and insensitive to variations in the Signal to Noise ratio and in the limiting surface brightness of the observations \citep{petr76,lotz04}. The Petrosian radius $r_{\rm p}$ is defined to be the radius at which the surface brightness $\mu(r_{\rm p}$) is equal to a factor of $\eta$ times the mean surface brightness $\bar{\mu}(r < r_{\rm p})$ within the $r_{\rm p}$. That is, $\mu(r_{\rm p}) \ = \ \eta \times \bar{\mu} \ (r < r_{\rm p})$. A constant $\eta$ value is typically set to 0.2 \citep[e.g.,][]{petr76,cons00,shim01,lotz04,guai15}. We adopt the same $\eta$ value of 0.2 for our size measurement.

To solve for the above Petrosian equation, we measure the surface brightness profile of galaxies by performing the \textsf{IRAF} ellipse task. To account for the PSF effect on the Petrosian radius measurement, we adopt the same PSF-deconvolved and sky-subtracted COS/NUV images that we processed in Paper I (see Section 2.2 for details). During the ellipse task, we fix the galaxy center and the ellipticity $\epsilon$ with the values obtained from the GALFIT results. The measured surface brightness profile of sample GPs is used to find the $r_{\rm p}$. The exact value of $r_{\rm p}$ is calculated by first finding the two closest radii to $r_{\rm p}$ at which the surface brightnesses are measured, and then performing linear interpolation or extrapolation with these two closest radii to derive the exact $r_{\rm p}$ that satisfies the Petrosian equation.

The typical difference between the exact value of $r_{\rm p}$ derived and the two closest radii found in the measured surface brightness profile is only 0.22 pixels (equivalently 0.0052 arcsec). This small interpolation/extrapolation corresponds to the $\sim 4 \%$ relative difference between the derived $r_{\rm p}$ and the two closest radii (that is, $1-(\bar{r_{\rm cl}}/r_{\rm p}) \sim 0.04$ where $\bar{r_{\rm cl}}$ is the mean radius of the two closest radii). Therefore, we do not expect the details of our interpolation procedure (e.g., the polynomial degree used) to significantly change our measured sizes.

We subsequently measure the UV-continuum luminosity $L_{\rm p}$ and the Petrosian half-light radius $r_{\rm 50}$ based on the $r_{\rm p}$ by following the conventional definitions of $L_{\rm p}$ and $r_{\rm 50}$ \citep[e.g.,][]{shim01}. The $L_{\rm p}$ is defined to be the total flux within the enclosed area of 2$r_{\rm p}$ from galaxy center, and the $r_{\rm 50}$ to be the radius within which the half of $L_{\rm p}$ is enclosed. As in the $r_{\rm p}$ measurement, we perform the same linear interpolation/extrapolation to derive the exact values of $L_{\rm p}$ and $r_{\rm 50}$. The measured $r_{\rm 50}$ of our sample GPs in pixel unit ranges from $2 \ {\rm pixel} \lesssim r_{\rm 50} \lesssim 17 \ {\rm pixel}$, with a median of 4.1 pixel, first and third quartile values of 2.6 pixel and 6.4 pixel, and a standard deviation of 3.1 pixel. The measurement uncertainties in $L_{\rm p}$ are derived considering the Poisson statistics (i.e., photon counting statistics) and propagation of the errors during the image calibration procedures such as flat-field correction obtained from the provided COS/NUV error images. For uncertainties in $r_{\rm 50}$, we adopted the derived fractional uncertainty of half-light radius based on median statistics and the associated aperture photometry uncertainty for LAEs at $z \sim 2-3$ in \cite{bond12} . The typical measurement uncertainties in $L_{\rm p}$ and $r_{\rm 50}$ in our sample GPs are 0.02 mag and 0.01 kpc, respectively.

Lastly, the measured $L_{\rm p}$ is corrected for galactic internal extinction ($A_{\rm int}$), Milky Way extinction ($A_{\rm MW}$), and the $k$-correction ($k$) with the values obtained from Paper I (see Section \ref{subsec:Sample selection} for details). By adopting the characteristic luminosity at $z \sim 3$ ($L_{\rm *, z=3}$) which corresponds to UV magnitude of -21, the $L_{\rm p}$ of our sample GPs ranges from 0.1$L_{\rm *, z=3}$ to 2.4$L_{\rm *, z=3}$, with a median $L_{\rm p}$ of 0.64$L_{\rm *, z=3}$. We also calculate the circularized half-light radius (i.e., \prad\ $= r_{\rm 50} \times \sqrt{b/a}$) and the circularized effecitve radius ($r_{\rm cir,eff}$ $= r_{\rm eff} \times \sqrt{b/a}$ where $= r_{\rm eff}$ is the effective radius along the major axis), respectively.  The measured $r_{\rm p}$, $r_{\rm 50}$, \prad , $r_{\rm cir,eff}$, \sersic\ index,  $b/a$, and $L_{\rm p}$ as well as the adopted extinction ($A_{\rm int}$ and $A_{\rm MW}$) and $k$-correction ($k$) values are listed in Table \ref{tab1}. 

\section{Results and Discussion}
\label{sec:results}

\subsection{Size Distribution of GPs}
\label{subsec:Size distribution}

\begin{figure*}[!htbp]
\centering
\includegraphics[width=1\textwidth]{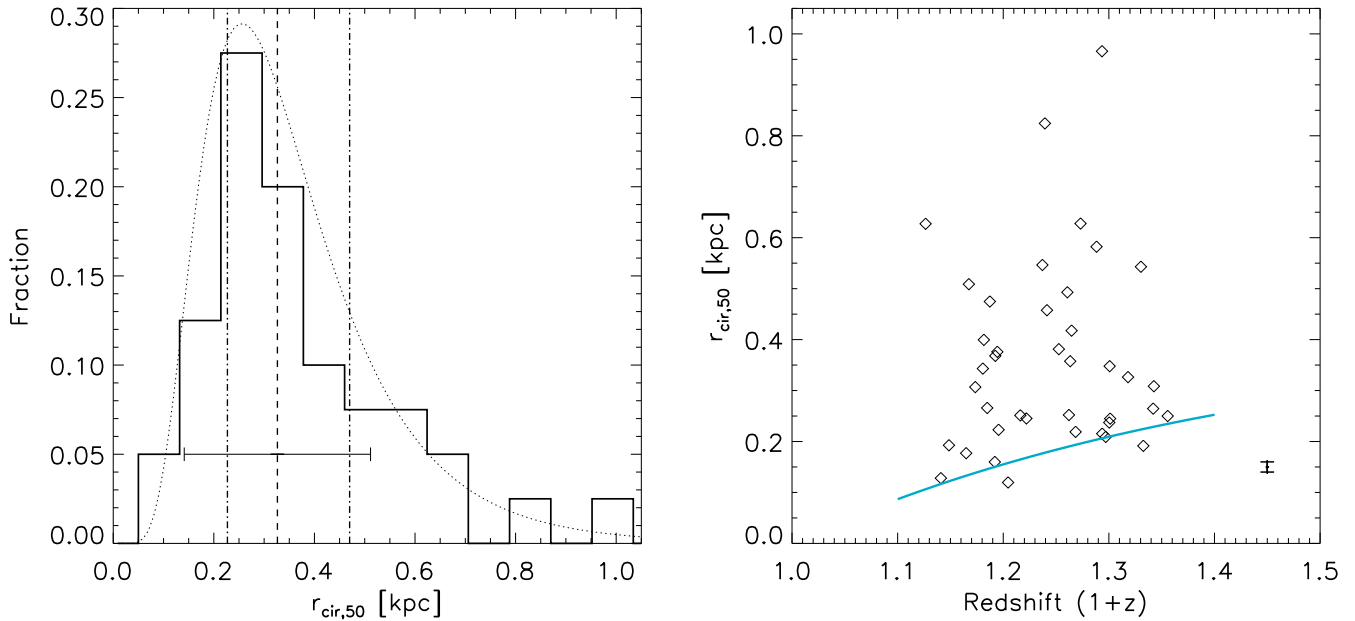}
\caption{Left: The size (\prad) distribution of 40 sample Green Pea galaxies. The vertical dashed line indicates the median \prad\ of 0.33 kpc, while the horizontal solid line indicates a $1\sigma$ of 0.19 kpc. The 25th and 75th percentiles of the size distribution are 0.23 kpc and 0.47 kpc, respectively, marked as the vertical dotted-dashed lines. The semi-interquartile range of the distribution is 0.12 kpc. The dotted curve shows the log-normal fit to the size distribution.  This fit has median size ($r_{\rm med}$) of 0.32 kpc and a width of ($\sigma_{\rm ln (r_{\rm cir,50})}$) of 0.48. The fitted log-normal probability density function is scaled by a factor of 0.1 to visualize with the actual distribution. Right: the size vs. redshift of GPs. The blue solid line shows the typical PSF FWHM of COS/NUV images in units of physical kpc. The panel shows that the sizes of sample GPs are mostly resolved with the NUV images. The typical measurement error in \prad\ is marked in the bottom right.}
\label{fig2}
\end{figure*}

In this section, we investigate the UV size distribution of our sample GPs. Figure \ref{fig2} shows the distributions of \prad\ (left panel) and of \prad\ vs. redshift (right panel), respectively. The left panel shows that our sample GPs show mostly small UV sizes with a median \prad\ of 0.33 kpc and a 1$\sigma$ spread of 0.19 kpc, marked as the vertical dashed line and the horizontal solid line, respectively. The size distribution is relatively narrow as its semi-interquartile range is only 0.13 kpc. The mean \prad\ of 0.36 kpc is larger than the median, due to the presence of a long tail toward large sizes as shown in the left panel of the figure\footnote{The typical size of our sample GPs (i.e., the median \prad\ of $0.33$ kpc)  corresponds to the angular size of $\sim$ 0.1 arcsec at the typical redshift of 0.25 of our sample GPs. We note that the measured small size of our sample GPs does not result from potential sample selection bias as our sample GPs were drawn, without imposing size criteria, from the parent sample Green Peas whose SDSS image has significantly larger seeing cutoff of $\gtrsim$ 1 arcsec that corresponds to $\sim$5 kpc at their typical redshifts (see Section 2.3 of \cite{card09} for further details)}. The right panel shows the size vs. redshift distribution. Compared to the typical PSF FWHM of COS/NUV images (i.e., the blue solid line), it shows that the sizes of sample GPs are mostly resolved with the NUV images.

We also fit the size distribution of our sample GPs with a log-normal distribution:
\begin{eqnarray}
p(r_{\rm cir,50}) = \frac{1}{\sigma_{\rm ln(r_{\rm cir,50})} r_{\rm cir,50} \sqrt{2 \pi}} \exp \Bigg[-\frac{{\rm ln}^{2}(r_{\rm cir,50}/r_{\rm med})}{2{\sigma_{\rm ln(r_{\rm cir,50})}}^{2}} \Bigg] \ ,
\label{Eq1}
\end{eqnarray}
where $p(r_{\rm cir,50})$ is the log-normal probability density function of \prad. Here $r_{\rm med}$ and $\sigma_{\rm ln(r_{\rm cir,50})}$ are the median size and the width of the fitted log-normal distribution, respectively. As shown in the left panel of the figure, the size distribution of our sample GPs is well fitted by a log-normal function (i.e., dotted line) with the associated $r_{\rm med} = 0.32\, \hbox{kpc}$ and $\sigma_{\rm ln(r_{\rm cir,50})} = 0.47$.

A relevant interpretation of the log-normal distribution of galaxy size might come from theoretical galaxy disk formation models \citep[e.g.,][]{peeb69,fall80,barn87,mo98} where the pre-collapse dark matter (DM) halo angular momentum is predicted to have a log-normal distribution. The associated dimensionless spin parameter $\lambda \equiv \frac{J{|E|}^{1/2}}{G{M}^{5/2}}$--- where $J$, $E$, $M$, and $G$ indicate the total angular momentum, energy, and mass of the system, and the gravitational constant, respectively--- follows the log-normal distribution with the associated peak value $\lambda_{\rm med}\sim 0.05$ and width $\sigma_{\rm ln(\lambda)}\sim 0.5-0.6$  \citep{barn87,warr92,cole96,davi09}. Assuming that the baryonic component acquired most of its angular momentum through its surrounding dark matter halos and that the galaxy disk size is largely determined by the multiplication of $\lambda$ by the virial radius ($R_{\rm vir}$) \citep[e.g.,][]{fall80,mo98}, it seems reasonable to expect that the galaxy disk size distribution also follows a log-normal distribution, reflecting the underlying dark matter halo properties. Our fitted log-normal UV size distribution of sample GPs with the associated $\sigma_{\rm ln(r_{\rm cir,50})}$ of 0.47 shows a $\sim 6\%$ narrower width compared to the predicted dark matter halo $\lambda$ distribution with the associated $\sigma_{\rm ln(\lambda)}$ of $\sim$ 0.5, but is broadly consistent with other types of star-forming galaxies at low and high redshifts \citep[e.g.,][]{dejo00,shen03,huan13}.

\subsection{UV Size-Luminosity Relation of GPs}
\label{subsec:Size Luminosity relation}
The size-luminosity relationship for galaxies is a key piece to understand the growth history of galaxies over cosmic time.  For the high-\textit{z} universe, the UV size-luminosity relations of star-forming galaxies such as LAEs and UV continuum-selected Lyman-break Galaxies (LBGs) have been previously investigated \citep[e.g.,][]{oesc10,huan13,jian13,ono13,holw15,holw20,shib15,brid19}. The size-luminosity relation is often parameterized as a power law with a slope of $\alpha$:
\begin{eqnarray}
r_{\rm cir,50} = r_{0}\Bigg(\frac{L_{\rm UV}}{L_{0}} \Bigg)^{\alpha} ,
\label{Eq2}
\end{eqnarray}
where $L_{\rm UV}$, $L_{0}$, and $r_{0}$ are the UV luminosity of galaxies, a fiducial UV luminosity (which we take as the same redshift 3 characteristic luminosity we used in section 2.2, i.e., $L_0 = L_{\rm *, z=3}$, corresponding to $M_{\rm UV} = -21$), and the size at $L_{0}$, respectively. 

The fitted $\alpha$ and $r_{0}$ of our sample GPs are 0.30 $\pm$ 0.09 and 0.38 $\pm$ 0.03 kpc, respectively. The slope value is consistent within the uncertainties with those for high-\textit{z} ($2 \lesssim z \lesssim 8$) star-forming galaxy populations, which show a slope range of $0.15 \lesssim \alpha \lesssim 0.5$ with the typical $\alpha$ of $\sim$ 0.27 \citep{graz12,jian13,huan13,shib15,curt16}. A brief interpretation on the slope of the size-luminosity relation might come from theoretical galaxy disk formation models (as in Section 3.1). Based on the Tully-Fisher relation and the assumed constant mass-to-light ($M/L$) ratio of disks, the fiducial slope of the size-luminosity derived is 1/3 \citep{mo98}. The deviation from the fiducial slope value could be due to varying $M/L$ ratio and/or stellar feedback such as SN feedback from star formation in disks, as such effects can change the slope value \citep{wyit11,liu17}.   The fitted slope of our sample GPs is broadly in agreement within the uncertainties with the suggested value of 0.25 from the semi-analytic model that accounts for the effect of SN feedback \citep{liu17}.

\begin{figure}
\centering
\includegraphics[width=0.5\textwidth]{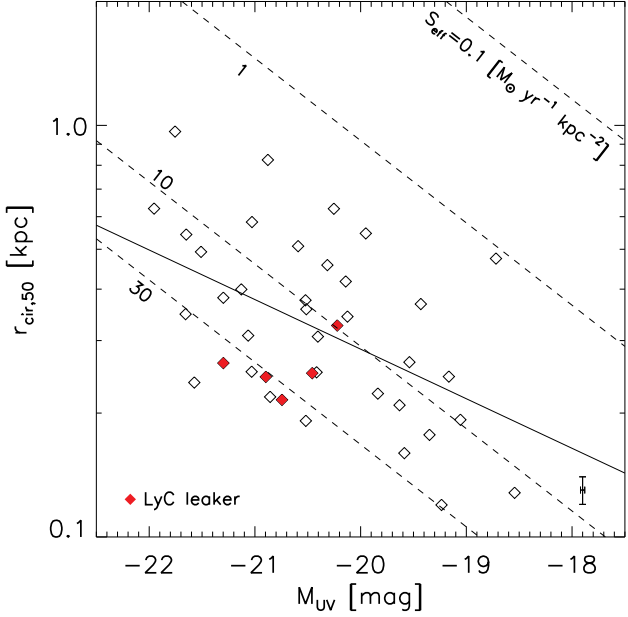}
\caption{The UV-continuum size vs. luminosity relation of GPs. The solid line shows the linear fit to the UV size-luminosity relation (i.e., Equation \ref{Eq2}), with the best-fit slope $\alpha$ of 0.30 $\pm$ 0.09 and the intercept $r_{\rm 0}$ of $0.38 \pm 0.03$ kpc at $M_{UV} -21$. The dashed lines indicate a constant effective star formation intensity (SFI, equivalent to star formation rate surface density) defined in Section \ref{subsec:Size Luminosity relation}. The comparison of individual data points with the constant effective SFI lines shows that GPs have SFI that is relatively high (by least 2 order of magnitude) compared to typical star-forming galaxies. (GPs have $S_{\rm eff} \gtrsim 1 M_{\sun} \ \rm{yr}^{-1} \ \rm{kpc^{-2}}$, vs.\ $\sim 0.01 M_{\sun} \ \rm{yr}^{-1} \ \rm{kpc^{-2}}$ for typical low-redshift star forming galaxies). The typical measurement errors of 0.01 kpc and 0.02 mag in \prad\ and UV magnitude are marked in bottom right. The red-filled diamonds are the five confirmed LyC leakers from \cite{izot16}.}
\label{fig3}
\end{figure}

Figure \ref{fig3} shows the UV size-luminosity relation of our sample GPs.   Comparing the distribution of individual data points to lines of constant effective star formation intensity ($S_{\rm eff}$) demonstrates that all of our sample GPs have  $S_{\rm eff} \gtrsim 1 \ M_{\sun} \ \rm{yr}^{-1} \ \rm{kpc^{-2}}$, which is 2+ orders of magnitude above those of typical star-forming galaxies ($\lesssim 0.01 \ M_{\sun} \ \rm{yr}^{-1} \ \rm{kpc^{-2}}$) \citep[e.g.,][]{kenn98}.   This is mainly due to the compact sizes of GPs given their UV luminosity.   Here, the effective SFI is defined as  $S_{\rm eff} \equiv \frac{\rm{SFR}}{2 \pi r_{\rm{cir,50}}^{2}}  \left (\frac{M_{\sun} \ \rm{yr}^{-1}}{{\rm kpc}^2} \right )$.  To convert the measured UV luminosity to the corresponding SFR, we adopt the solar bolometric magnitude of 4.74 \citep{bess98}, and the UV to bolometric luminosity ($L_{\rm bol}$) ratio ($L_{\rm UV}/L_{\rm bol}$) of 0.33 and the scale factor $L_{\rm bol}/(4.5 \times 10^{9} L_{\sun}) = {\rm SFR}/(1 M_{\sun} \ \rm{yr}^{-1})$ that are derived from the starburst population modeling by \cite{meur97}. The starburst population modeling assumes a solar metallicity and the Salpeter initial mass function \citep{salp55} with lower and upper mass limits of 0.1 $M_{\sun}$ and 100 $M_{\sun}$, respectively.

The effective SFI of our sample GPs mostly ranges from 1 to $\sim$ 30 $M_{\sun} \ \rm{yr}^{-1} \ \rm{kpc^{-2}}$ as shown in the figure. The range of the effective SFI is similar to the central 250 pc region SFI ($S_{\rm 250 pc}$) that we reported in Paper I, which ranges from 2.3 to 46 $M_{\sun} \ \rm{yr}^{-1} \ \rm{kpc^{-2}}$ with a mean $S_{\rm 250 pc}$ of 15 $M_{\sun} \ \rm{yr}^{-1} \ \rm{kpc^{-2}}$ for the same sample GPs. This is not unexpected, since the typical \prad\ of our sample GPs is $\sim$ 0.3 kpc (i.e., Figure \ref{fig2}), so that the effective SFI is measured in a region that largely overlaps the central 250pc for most sample objects.

\subsection{Correlations between UV Size and the \lya\ Properties of GPs}
\label{subsec:Correlations between UV size and LyA properties}

\begin{figure}
\centering
\includegraphics[width=0.5\textwidth]{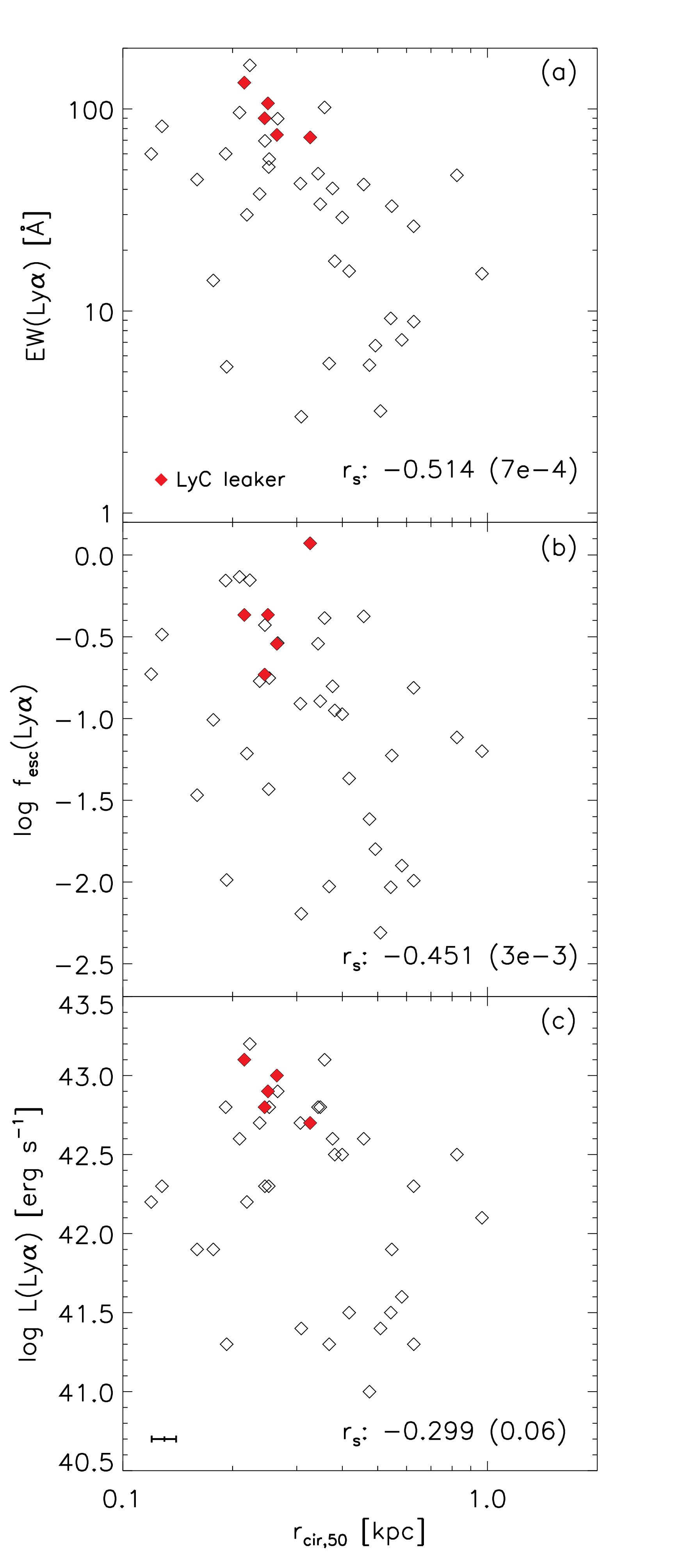}
\caption{The correlations between \prad\ vs. \ewlya\ (panel (a)), \fesclya\ (panel (b)), and \lyalum\ (panel (c)). The format is the same as in Figure \ref{fig3}. The associated $r_{\rm{s}}$ values ($p$-values) are shown in the bottom right of each panel. Note that the presented \lya\ properties (i.e., \ewlya, \fesclya, and \lyalum) largely show anticorrelations with \prad\ in our sample GPs, suggesting that small sizes are preferred for significant \lya\ emission. The horizontal error bar in panel (c) indicates the typical measurement uncertainties in \prad . The typical measurement uncertainties of the adopted \ewlya, \fesclya, and \lyalum\ from Y17 are $\sim 15 \%$ mainly dominated by the systematic error.}
\label{fig4}
\end{figure}

In this section, we investigate the possible correlations between UV size and the \lya\ properties such as \ewlya, \fesclya, and \lyalum\ in our sample GPs. Figure \ref{fig4} shows the correlations between \prad\ and \ewlya, \fesclya, and \lyalum. Panel (a) shows the correlation between \prad\ and \ewlya. Based on the distribution and the associated Spearman correlation coefficient (hereafter, $r_{\rm s}$) ($p$-value) of -0.514 (7 $\ \times \  10^{-4}$), there is a statistically significant anticorrelation between the two parameters such that a smaller UV size is preferred for a larger \ewlya.

A similar anticorrelation with the associated $r_{\rm s}$ ($p$-value) of $-0.64$ (0.03) was reported based on 12 local star-forming galaxies ($0.03 < z < 0.2$) from the Lyman alpha reference sample (LARS) \citep{guai15}. For high-\textit{z} LAEs at $z \simeq 2.1$ and 3.1, there is a systematic trend seen such that high \ewlya\ sample LAEs have a smaller median UV size compared to that of low \ewlya\ sample LAEs \citep[i.e.,][]{bond12}. Also, a linear fit to the UV size and \ewlya\ relation for a sample of high-\textit{z} LAEs at $2 \lesssim z \lesssim 6$ showing a negative slope of -3.5 by \cite{paul18} seems qualitatively consistent with the anticorrelation shown in our sample GPs.

A correlation between \prad\ and \fesclya\ is shown in panel (b), which shows a qualitatively similar anticorrelation to that in panel (a).  The associated $r_{\rm s}$ ($p$-value) is -0.451 (3 $\times \ 10^{-3}$). The distribution and the associated $r_{\rm s}$ seem to suggest that a smaller UV size is preferred for larger \fesclya . The same correlation was investigated based on the 12 LARS sample galaxies, although their $r_{\rm s}$ ($p$-value) of $-0.22$ (0.48) is not statistically significant \citep{guai15}.

Panel (c) shows the correlation between \prad\ and \lyalum. First, we note that one galaxy (GP ID 0339-0725) is omitted from this panel only, since it has no reported \lyalum\ in Y17. Compared to the other correlations examined above with \ewlya\ and \fesclya , \lyalum\ does not show a statistically significant anticorrelation with \prad , as the associated $r_{\rm s}$ ($p$-value) is only -0.299 (0.06). Nonetheless, it also seems evident that the large UV size is not preferred for high \lyalum\ , as a previous study \citep[][]{paul18} also shows a negative slope $-0.51$ of the linear fit to the UV size and \lyalum\ relation based on a sample of high-\textit{z} LAEs.

The five confirmed LyC leakers in our sample GPs are marked as the red diamonds in Figure \ref{fig4}. Due to their high values of \ewlya , \fesclya , and \lyalum\ (Y17), and relatively small \prad\ compared to the sizes of the entire sample GPs, the confirmed LyC leakers occupy the top left of each panel (that is, small \prad\ and large \ewlya\ (panel (a)), \fesclya\ (panel (b)), and \lyalum\ (panel (c)), respectively).

\subsection{The Size of GPs compared to high-$z$ LAEs}
\label{subsec:size comparison with high-z LAEs}
Malhotra et al. (2012) show that the sizes of \lya\ galaxies are similar at all redshifts, so GPs should have similar sizes to high-\textit{z} galaxies. Shibuya et al. 2019, however, argue that the sizes of \lya\ galaxies evolve with redshift as $r_{\rm  cir, eff} \propto (1+z)^{\beta}$ with ($\beta$) values of $-1.37$, similar to those of Lyman Break Galaxies (LBGs). At the redshift of GPs ($ z =0.1-0.3$) these predictions differ by a factor of 10. But the typical size (i.e., the median $r_{\rm cir, eff}$) of \lya\ galaxies seems to {\it decrease} by a factor of 1.6 instead of increasing by a factor of 10 from $z > $ 2 to $z \simeq$ 0.3.

\begin{figure*}[!htbp]
\centering
\includegraphics[width=1.0\textwidth]{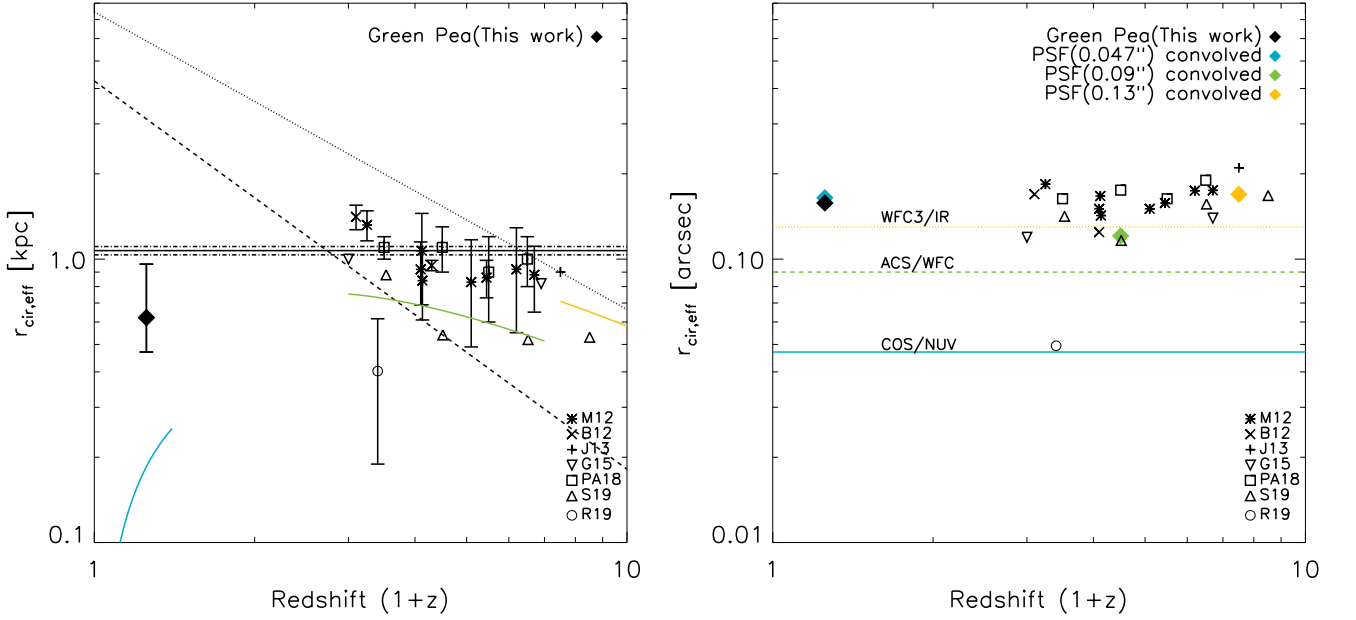}
\caption{Left: the median UV-continuum size (as measured by $r_{\rm cir, eff}$ in the same way at all redshifts)  of sample GPs (the filled diamond at $1+z \sim 1.25$) compared to those of high-\textit{z} LAEs. The sizes of high-\textit{z} LAEs are from the literature  (asterisk: Malhotra et al. (2012; M12), x mark: Bond et al. (2012; B12), cross: Jiang et al. (2013; J13), downward triangle: Guaita et al. (2015; G15), square: Paulino-Afonso et al. (2018; PA18), triangle: Shibuya et al. (2019; S19), and circle: Ritondale et al. (2019; R19)). The associated error bar indicates the 1$\sigma$ standard deviation of size distribution of respective sample galaxies except for that of sample GPs and of R19 which show the associated interquartile range and the uncertainties derived from random MCMC realizations to quantify errors of their size measurements of lensed LAEs. The data points for B12 for the $1+z=4.1$ sample LAEs and for G15 for the $1+z=3.1$ sample LAEs are shifted by 0.2 in x-axis for clarity. The solid horizontal line is the weighted mean size of the combined sample LAEs of M12 and PA18, and the dotted-dashed lines are the associated weighted standard deviation of the weighted mean size. The colored solid lines are the physical sizes corresponding to the typical PSF FWHM of different instruments that are marked within their relevant redshift range for LAE size measurements as indicated in the right panel. The dashed line is the predicted size evolution of LAEs based on S19. The dotted line is the predicted size evolution of LBGs from \cite{bouw04}. Our sample GPs do not follow either of these predicted size evolution models. The small median size of sample GPs does not change significantly even if the sample GPs are divided into specific $L_{\rm UV}$ and \ewlya\ bins as listed in Table 2. See the text for further details. Right: same as the left panel, but for the corresponding angular sizes of GPs and high-\textit{z} LAEs that are PSF-convolved in quadrature, except for the sizes of lensed LAEs (R19, circle) and of LAEs from B12 and G15. The typical PSF FWHM of different \textit{HST} instrument filters are marked as the colored horizontal lines. The blue, green, and orange diamonds are the PSF-convolved angular sizes of original median physical size of sample GPs (black diamond) by employing the different PSF FWHM indicated by each corresponding color. Note that the PSF-convolved median sizes of sample GPs and high-\textit{z} LAEs become 
undersampled as we go to higher redshift. See the text for further details. The non-evolving size of GPs suggests that a small size is crucial for a galaxy to become a Ly$\alpha$-emitter.}
\label{fig5_6}
\end{figure*}

The left panel of Figure \ref{fig5_6} shows the compilation of the median physical sizes of our sample GPs and high-\textit{z} LAEs as a function of redshift. For the sizes of 
high-\textit{z} LAEs at $2 \lesssim z \lesssim 7$, we refer to previous studies \citep[i.e.,][]{malh12,bond12,jian13,guai15,paul18,shib19,rito19}. The sizes from \cite{guai15} are based on their simulated images, which were produced by taking low-redshift LARS-LAE galaxies and simulating the effects of dimming due to redshift and observational effects at $z \sim$ 2 and 5.7, respectively. The median size of our sample GPs (i.e., the filled diamond) does not follow the predicted size growth evolution of either LAEs (i.e., the dashed line) or LBGs (i.e., the dotted line) as suggested by previous studies (i.e., Bouwens et al. 2004 for LBGs and Shibuya et al. 2019 for LAEs)---  the plotted size evolution curves of LAEs and LBGs in the figure show the reported power-law fitting slope ($\beta$) values of $-1.37 \pm 0.65$ and $-1.05 \pm 0.21$ in the relation of $r_{\rm  cir, eff} \propto (1+z)^{\beta}$ for high-\textit{z} LAEs and LBGs, respectively. Also, to derive the intercept of each size evolution curve we use the reported size at $z \sim 3.5$ (for LAEs) and at $z \sim 5$ (for LBGs), respectively, from the previous studies (see Figure 9 of \cite{shib19} for LAEs and Figure 4 of \cite{bouw04} for LBGs for details).

In particular, the median size (i.e., the median effective radius $r_{\rm  cir, eff}$) of 0.62 kpc of our sample GPs is approximately a factor of 7 smaller than predicted using the LAE size evolution models from \citet{shib19}. Our size measurement uncertainties are typically $\sim$ 0.03 kpc and cannot explain the discrepancy. 
This size difference does not change significantly even if we divided our sample GPs depending on the specific $L_{\rm UV}$ and/or \ewlya\ bins similar to previous studies for the size of LAEs.  For example, a subset of our sample with luminosities $0.3 < L / L_{\rm *, z=3} < 1$ and \ewlya\ $> 20\angstrom$ contains 16 GPs whose median size is $0.64 ^{+0.27}_{-0.16}$ kpc, indistinguishable from the result for our full sample.  The mean and median sizes of our GP subsamples with various different $L_{\rm UV}$ and \ewlya\ cuts are listed in Table \ref{tab2}.

On average GPs  show radii that are a factor of $\sim 1.6$ smaller compared to high-z LAEs (i.e., 0.62 kpc vs. $\sim$ 1 kpc). This difference is due to both the angular resolution of the \textit{HST} instruments which would increase as (1+$z$) for the same rest wavelength, and also due to the angular diameter distance increase with redshift. The spatial resolution for different instruments is shown in the right panel of Figure \ref{fig5_6}. If we assume that the sizes of LAEs is unchanged with redshift and convolve the median GPs radius of 0.62 kpc with the typical PSF FWHM of 0.09$''$ of ACS/WFC instrument or the typical PSF FWHM of 0.13$''$ of WFC3/IR, the measured radii of 
high-\textit{z} LAEs are consistent with the radii of GPs.  In the right panel of Figure \ref{fig5_6} we show the results of convolving measured median angular size of GPs (black diamond) with three different PSF FWHM of different instruments (that is, COS/NUV (blue diamond), ACS/WFC (green diamond), and WFC3/IR (orange diamond)) to show the effects of different PSF FWHM on the measured size of galaxies. Moreover, the PSF-convolved median sizes of GPs and high-\textit{z} LAEs with ACS and WFC3 instruments in the unit of pixels show $\sim7.1$ pixels, $\sim 2.4$ pixels, and $\sim 1.3$ pixels, respectively. This suggests that the measured median sizes of galaxies become getting undersampled as we go to higher redshift.

This result is supported by measurements of the 
UV continuum size for lensed LAEs at $z=2$--$3$ by \citet{rito19},
who find  $r_{\rm maj} = 0.561_{-0.110}^{+0.013}$ kpc, with the corresponding circularized size of $0.403 \pm 0.213$ kpc (the open circle in the left panel of Figure \ref{fig5_6}).   This measurement is $\sim 2.5$ times smaller than the sizes measured for non-lensed LAEs at similar redshifts, and broadly comparable to the sizes of GPs, suggesting again that spatial resolution limits size measurements for high-redshift LAEs.

\section{Summary and Conclusions}
\label{sec:Summary and Conclusions}
In this study, we have measured the UV-continuum size and luminosity of a sample of 40 GPs. As GPs are the best local analogs to high-\textit{z} LAEs, their physical proximity and the high spatial resolution of COS/NUV images (PSF FWHM of 0.047 arcsec with high sampling of $0.0235$ arcsec $ \rm{pixel}^{-1}$) have enabled us to study the spatially resolved UV size and luminosity properties of LAEs in local universe and compare with those of high-\textit{z} LAEs in detail. The main results are summarized as follows.

\begin{itemize}
\item GPs show compact sizes with a median \prad\ of 0.33 kpc. The size distribution (Figure \ref{fig2}) is narrow (with its semi-interquartile range of 0.12 kpc) and is well fitted by a log-normal distribution with the associated width $\sigma_{\rm ln(r_{\rm cir,50})}$ of 0.47. While the log-normal shape of the size distribution of GPs is largely consistent with continuum-selected star-forming galaxies, the peak (\prad \ at $\simeq 0.3$ kpc) of the distribution is smaller due to the compact sizes of GPs.

\item The UV size-luminosity relation of GPs (Figure \ref{fig3}) shows a fitted slope of $0.30 \pm 0.09$ and an intercept $r_{\rm 0}$ at $M_{\rm UV} = -21$ of 0.38 $\pm 0.03$ kpc. The fitted slope value is consistent with other types of star-forming galaxies at low and high redshifts within the uncertainties. ($0.15 \lesssim \alpha \lesssim 0.35$). However, the effective SFI (i.e., star formation rate surface density) of our sample GPs shown in the size-luminosity relation indicates relatively high SFI for GPs (1$-$30 $M_{\sun} \ \rm{yr}^{-1} \ \rm{kpc^{-2}}$),  exceeding the typical SFI of star-forming galaxies by 2 orders of magnitude or more.

\item There are anticorrelations between UV-continuum size (\prad), \ewlya , \fesclya , and \lyalum\ in sample GPs (Figure \ref{fig4}). In particular, the anticorrelations between \prad, \ewlya, and \fesclya are statistically significant, and suggest that small UV-continuum sizes are associated with \lya\ galaxies.

\item The size comparison of sample GPs with high-\textit{z} LAEs (Figure \ref{fig5_6}) shows that the typical size of GPs (i.e., the median $r_{\rm  cir, eff}$ of 0.62 kpc) is a factor of $\sim 1.6$ smaller than those of high-\textit{z} LAEs ($\simeq 1$ kpc) at $2 \lesssim z \lesssim 6$. Also, the compact size of GPs does not seem to follow the predicted size growth with time of either high-\textit{z} LAEs or LBGs, rather showing a factor of $\sim 7$ smaller size than the predicted size at low redshifts. The smaller size of our sample GPs compared to those of high-\textit{z} LAEs, is, however, due to the different spatial resolution between low-\textit{z} GPs and high-\textit{z} LAEs (Figure 4). Once the different resolution is properly taken into account, their sizes are consistent with negligible evolution.

\end{itemize}

All in all, our UV size and luminosity analysis of Green Pea galaxies suggests that a compact/small size is crucial for escape of \lya\ photons, and that LAEs show constant characteristic size independent of their redshift. Also, our study implies that small sizes can help select Ly$\alpha$-emitters.

\begin{table*}[ht]
\centering
\fontsize{6.}{8.}\selectfont
\setlength{\tabcolsep}{2.pt}
\begin{threeparttable}
\caption{The Measured Parameters of Sample Galaxies}
\begin{tabular}{lllllllllllllll}
\hline \hline
Green Pea ID\tnote{a} & SDSS ObjID\tnote{b} & $ m_{\rm UV}$ \tnote{c,d} & $ M_{\rm UV}$ \tnote{e,d} & \prad \tnote{f,g}  & \prad \tnote{h,g} & $r_{\rm cir,p}$ \tnote{i,g} & $r_{\rm cir,p}$ \tnote{j,g} & {$ r_{\rm cir,eff}$}\tnote{k} & {$ r_{\rm cir,eff}$}\tnote{l} & n \tnote{m} & $ b/a$ \tnote{n} & $A_{\rm int}$ \tnote{o} & $A_{\rm MW}$ \tnote{p} & $k$ \tnote{q} \\
& & (mag) & (mag) & (arcsec) & (kpc) & (arcsec) & (kpc) & (arcsec) & (kpc) & & & (mag) & (mag) & (mag) \\
\hline

0055-0021 & 1237663783666581565 &   19.85 $\pm$   0.03 &  -20.59 $\pm$  0.03 &  0.233 $\pm$  0.006 &  0.509 $\pm$  0.016 &  0.526 $\pm$ 0.013 &  1.147 $\pm$ 0.037 &  0.152 $\pm$ 0.009 &  0.435 $\pm$ 0.019 &  1.00 $\pm$ 0.18 &  0.58 $\pm$ 0.07  & 0.85 & 0.184 & 0.123 \\
0303-0759 & 1237652900231053501 &  20.73 $\pm$  0.02 &  -19.35 $\pm$ 0.02 &  0.072 $\pm$ 0.001 &   0.177 $\pm$ 0.003 &  0.173 $\pm$ 0.002 &  0.424 $\pm$ 0.006 &  0.212 $\pm$ 0.010 &  0.599 $\pm$ 0.025 &  8.00 $\pm$ 0.25 &  0.75 $\pm$ 0.01 & 0 & 0.716 & 0.128 \\
0339-0725 & 1237649961383493869 &  19.97 $\pm$ 0.02 &  -21.51 $\pm$ 0.02 &  0.149 $\pm$ 0.002 &   0.493 $\pm$ 0.007 &   0.289 $\pm$ 0.004 &   0.954 $\pm$ 0.014 &    0.213 $\pm$ 0.006 &   0.859 $\pm$ 0.020 &    3.38 $\pm$ 0.10 &   0.67 $\pm$ 0.01 & 0.387 & 0.447 & -0.043 \\
0749+3337 & 1237674366992646574 &   20.07 $\pm$ 0.02 &   -21.95 $\pm$ 0.02 &   0.227 $\pm$ 0.003 &   0.628 $\pm$ 0.011 &   0.345 $\pm$ 0.004 &   0.954 $\pm$ 0.017 &   0.308 $\pm$ 0.011 &   1.284 $\pm$ 0.029 &   3.52 $\pm$ 0.07 &   0.44 $\pm$ 0.01 & 0.832 & 0.405 & -0.065 \\
0751+1638 & 1237673807042708368 &   21.43 $\pm$ 0.03 &   -20.14 $\pm$ 0.03 &   0.127 $\pm$ 0.003 &   0.418 $\pm$ 0.012 &   0.262 $\pm$ 0.006 &   0.862 $\pm$ 0.024 &   0.383 $\pm$ 0.055 &   1.560 $\pm$ 0.182 &   6.84 $\pm$ 0.50 &   0.65 $\pm$ 0.02 & 0.609 & 0.264 & -0.050 \\
0805+0925 & 1237667729656905788 &   21.55 $\pm$ 0.03 &   -21.65 $\pm$ 0.03 &   0.128 $\pm$ 0.003 &   0.543 $\pm$ 0.014 &   0.247 $\pm$ 0.006 &   1.044 $\pm$ 0.027 &  0.315 $\pm$ 0.037 &   1.498 $\pm$ 0.156 &   5.20 $\pm$ 0.39 &   0.79 $\pm$ 0.03 & 1.686 & 0.154 & -0.160 \\
0815+2156 & 1237664668421849521 &   20.75 $\pm$ 0.01 &   -18.54 $\pm$ 0.01 &   0.060 $\pm$ 0.001 &   0.128 $\pm$ 0.002 &   0.138 $\pm$ 0.001 &   0.294 $\pm$ 0.003 &  0.205 $\pm$ 0.010 &   0.510 $\pm$ 0.021 &   8.00 $\pm$ 0.26 &   0.74 $\pm$ 0.01 & 0.054 & 0.298 & 0.173 \\
0822+2241 & 1237664092897083648 &    20.80 $\pm$ 0.02 &   -20.42 $\pm$ 0.02 &   0.076 $\pm$ 0.002 &   0.251 $\pm$ 0.005 &   0.133 $\pm$ 0.003 &   0.437 $\pm$ 0.008 &  0.284 $\pm$ 0.024 &   0.997 $\pm$ 0.08 &   8.00 $\pm$ 0.44 &   0.88 $\pm$ 0.02 & 0.781 & 0.329 & 0.035 \\
0911+1831 & 1237667429018697946 &   20.53 $\pm$ 0.01 &   -21.03 $\pm$ 0.01 &   0.074 $\pm$ 0.007 &   0.252 $\pm$ 0.003 &   0.146 $\pm$ 0.002 &   0.494 $\pm$ 0.006 &  0.136 $\pm$ 0.004 &   0.552 $\pm$ 0.014 &   8.00 $\pm$ 0.23 &   0.70 $\pm$ 0.01 & 0.686 & 0.206 & -0.046 \\
0917+3152 & 1237661382232768711 &   20.42 $\pm$ 0.02 &   -21.57 $\pm$ 0.02 &   0.057 $\pm$ 0.001 &   0.237 $\pm$ 0.004 &   0.108 $\pm$ 0.002 &   0.452 $\pm$ 0.007 &  0.121 $\pm$ 0.005 &   0.542 $\pm$ 0.023 &   8.00 $\pm$ 0.45 &   0.88 $\pm$ 0.02 & 0.783 & 0.143 & -0.111 \\
0925+1403\tnote{r} & 1237671262812897597 &   20.96 $\pm$ 0.01 &   -20.90 $\pm$ 0.01 &   0.058 $\pm$ 0.001 &   0.245 $\pm$ 0.002 &   0.122 $\pm$ 0.001 &   0.514 $\pm$ 0.005 &  0.121 $\pm$ 0.004 &   0.539 $\pm$ 0.019 &   8.00 $\pm$ 0.33 &   0.89 $\pm$ 0.02 & 0.555 & 0.225 & -0.112 \\
0926+4428 & 1237657630590107652 &   19.91 $\pm$ 0.03 &   -20.12 $\pm$ 0.03 &   0.137 $\pm$ 0.003 &   0.343 $\pm$ 0.009 &   0.263 $\pm$ 0.006 &   0.660 $\pm$ 0.018 &  0.155 $\pm$ 0.002 &   0.472 $\pm$ 0.004 &   4.09 $\pm$ 0.09 &   0.68 $\pm$ 0.01 & 0.292 & 0.132 & 0.099 \\
0927+1740 & 1237667536393142625 &   20.88 $\pm$ 0.02 &   -21.03 $\pm$ 0.02 &   0.149 $\pm$ 0.003 &   0.582 $\pm$ 0.011 &   0.322 $\pm$ 0.006 &   1.256 $\pm$ 0.024 &  0.421 $\pm$ 0.036 &   1.824 $\pm$ 0.139 &   4.84 $\pm$ 0.26 &   0.81 $\pm$ 0.02 & 0.742 & 0.220 & -0.090 \\
1009+2916 & 1237665126921011548 &   21.18 $\pm$ 0.03 &   -19.17 $\pm$ 0.03 &   0.078 $\pm$ 0.002 &   0.245 $\pm$ 0.006 &   0.150 $\pm$ 0.003 &   0.474 $\pm$ 0.011 &  0.222 $\pm$ 0.026 &   0.794 $\pm$ 0.083 &   8.00 $\pm$ 0.61 &   0.78 $\pm$ 0.03 & 0 & 0.163 & 0.024 \\
1018+4106 & 1237661851459584247 &   20.90 $\pm$ 0.02 &   -19.95 $\pm$ 0.02 &   0.164 $\pm$ 0.003 &   0.546 $\pm$ 0.011 &   0.276 $\pm$ 0.005 &   0.923 $\pm$ 0.018 &  0.326 $\pm$ 0.037 &   1.224 $\pm$ 0.122 &   7.42 $\pm$ 0.49 &   0.79 $\pm$ 0.03 & 0.380 & 0.099 & -0.002 \\
1025+3622 & 1237664668435677291 &   18.83 $\pm$ 0.01 &   -20.25 $\pm$ 0.01 &   0.400 $\pm$ 0.003 &   0.627 $\pm$ 0.007 &   0.878 $\pm$ 0.007 &   1.376 $\pm$ 0.015 &  0.257 $\pm$ 0.008 &   0.582 $\pm$ 0.013 &   4.82 $\pm$ 0.09 &   0.48 $\pm$ 0.01 & 0.338 & 0.082 & 0.201 \\
1032+2717 & 1237667211592794251 &   20.89 $\pm$ 0.02 &   -19.43 $\pm$ 0.02 &   0.120 $\pm$ 0.002 &   0.368 $\pm$ 0.007 &   0.261 $\pm$ 0.005 &   0.802 $\pm$ 0.015 &  0.371 $\pm$ 0.028 &   1.189 $\pm$ 0.085 &   5.51 $\pm$ 0.28 &   0.92 $\pm$ 0.02 & 0.384 & 0.150 & 0.077 \\
1054+5238 & 1237658801495474207 &   19.65 $\pm$ 0.01 &   -21.30 $\pm$ 0.01 &   0.100 $\pm$ 0.001 &   0.381 $\pm$ 0.004 &   0.204 $\pm$ 0.002 &   0.775 $\pm$ 0.008 &  0.126 $\pm$ 0.002 &   0.497 $\pm$ 0.008 &   5.90 $\pm$ 0.14 &   0.93 $\pm$ 0.01 & 0.280 & 0.107 & -0.029 \\
1122+6154 & 1237655464839479591 &   21.29 $\pm$ 0.03 &   -19.23 $\pm$ 0.03 &   0.036 $\pm$ 0.001 &   0.120 $\pm$ 0.003 &   0.082 $\pm$ 0.002 &   0.276 $\pm$ 0.006 &  0.174 $\pm$ 0.015 &   0.583 $\pm$ 0.051 &   8.00 $\pm$ 0.61 &   1.00 $\pm$ 0.03 & 0.514 & 0.058 & 0.056 \\
1133+6514 & 1237651067351073064 &   20.35 $\pm$ 0.01 &   -20.31 $\pm$ 0.01 &   0.140 $\pm$ 0.001 &   0.458 $\pm$ 0.005 &   0.260 $\pm$ 0.002 &   0.852 $\pm$ 0.008 &  0.272 $\pm$ 0.008 &   1.035 $\pm$ 0.028 &   4.55 $\pm$ 0.10 &   0.74 $\pm$ 0.01 & 0.162 & 0.079 & -0.010 \\
1137+3524 & 1237665129613885585 &   19.59 $\pm$ 0.01 &   -20.52 $\pm$ 0.01 &   0.168 $\pm$ 0.002 &   0.376 $\pm$ 0.005 &   0.365 $\pm$ 0.003 &   0.817 $\pm$ 0.010 &  0.155 $\pm$ 0.004 &   0.502 $\pm$ 0.008 &   4.25 $\pm$ 0.07 &   0.48 $\pm$ 0.00 & 0.171 & 0.132 & 0.074 \\
1152+3400\tnote{r} & 1237665127467647162 &   20.79 $\pm$ 0.02 &   -21.30 $\pm$ 0.02 &   0.061 $\pm$ 0.001 &   0.265 $\pm$ 0.003 &   0.119 $\pm$ 0.002 &   0.513 $\pm$ 0.006 &  0.106 $\pm$ 0.003 &   0.515 $\pm$ 0.013 &   6.73 $\pm$ 0.26 &   0.79 $\pm$ 0.02 & 0.480 & 0.143 & -0.179 \\
1205+2620 & 1237667321644908846 &   21.29 $\pm$ 0.03 &   -21.06 $\pm$ 0.03 &   0.079 $\pm$ 0.002 &   0.309 $\pm$ 0.008 &   0.175 $\pm$ 0.004 &   0.682 $\pm$ 0.018 &  0.129 $\pm$ 0.009 &   0.628 $\pm$ 0.034 &   5.45 $\pm$ 0.30 &   0.64 $\pm$ 0.02 & 0.750 & 0.138 & -0.180 \\
1219+1526 & 1237661070336852109 &   20.18 $\pm$ 0.01 &   -19.84 $\pm$ 0.01 &   0.071 $\pm$ 0.007 &   0.223 $\pm$ 0.003 &   0.159 $\pm$ 0.002 &   0.496 $\pm$ 0.005 &  0.073 $\pm$ 0.001 &   0.238 $\pm$ 0.004 &   7.01 $\pm$ 0.20 &   0.93 $\pm$ 0.01 & 0 & 0.190 & 0.072 \\
1244+0216 & 1237671266571387104 &   19.95 $\pm$ 0.01 &   -20.88 $\pm$ 0.01 &   0.258 $\pm$ 0.002 &   0.824 $\pm$ 0.007 &   0.492 $\pm$ 0.004 &   1.570 $\pm$ 0.012 &  0.270 $\pm$ 0.005 &   1.021 $\pm$ 0.016 &   2.99 $\pm$ 0.04 &   0.71 $\pm$ 0.01 & 0.251 & 0.179 & -0.006 \\
1249+1234 & 1237661817096962164 &   20.73 $\pm$ 0.01 &   -20.51 $\pm$ 0.01 &   0.104 $\pm$ 0.008 &   0.358 $\pm$ 0.004 &   0.212 $\pm$ 0.002 &   0.726 $\pm$ 0.007 &  0.223 $\pm$ 0.008 &   0.906 $\pm$ 0.028 &   5.95 $\pm$ 0.15 &   0.71 $\pm$ 0.01 & 0.343 & 0.217 & -0.048 \\
1333+6246\tnote{r} & 1237651249891967264 &   21.16 $\pm$ 0.01 &   -20.22 $\pm$ 0.01 &   0.076 $\pm$ 0.001 &   0.326 $\pm$ 0.003 &   0.193 $\pm$ 0.002 &   0.830 $\pm$ 0.008 &  0.151 $\pm$ 0.005 &   0.702 $\pm$ 0.023 &   6.21 $\pm$ 0.20 &   0.86 $\pm$ 0.01 & 0 & 0.140 & -0.140 \\
1339+1516 & 1237664292084318332 &   20.86 $\pm$ 0.03 &   -19.59 $\pm$ 0.03 &   0.052 $\pm$ 0.001 &   0.160 $\pm$ 0.004 &   0.109 $\pm$ 0.003 &   0.332 $\pm$ 0.008 &  0.143 $\pm$ 0.007 &   0.458 $\pm$ 0.022 &   8.00 $\pm$ 0.43 &   0.91 $\pm$ 0.02 & 0.452 & 0.219 & 0.078 \\
1424+4217 & 1237661360765730849 &   20.32 $\pm$ 0.02 &   -19.54 $\pm$ 0.02 &   0.097 $\pm$ 0.002 &   0.266 $\pm$ 0.004 &   0.230 $\pm$ 0.003 &   0.630 $\pm$ 0.009 &  0.137 $\pm$ 0.005 &   0.424 $\pm$ 0.013 &   8.00 $\pm$ 0.28 &   0.78 $\pm$ 0.01 & 0.111 & 0.074 & 0.091 \\
1428+1653 & 1237668297680683015 &   19.33 $\pm$ 0.01 &   -21.13 $\pm$ 0.01 &   0.150 $\pm$ 0.002 &   0.400 $\pm$ 0.004 &   0.321 $\pm$ 0.003 &   0.856 $\pm$ 0.009 &  0.199 $\pm$ 0.004 &   0.608 $\pm$ 0.011 &   3.95 $\pm$ 0.07 &   0.76 $\pm$ 0.01 & 0.690 & 0.142 & 0.097 \\
1429+0643 & 1237662268069511204 &   19.49 $\pm$ 0.02 &   -20.40 $\pm$ 0.02 &   0.119 $\pm$ 0.002 &   0.307 $\pm$ 0.005 &   0.234 $\pm$ 0.003 &   0.606 $\pm$ 0.009 &  0.124 $\pm$ 0.002 &   0.365 $\pm$ 0.004 &   7.09 $\pm$ 0.13 &   0.77 $\pm$ 0.01 & 0.208 & 0.183 & 0.112 \\
1440+4619 & 1237662301362978958 &   20.13 $\pm$ 0.02 &   -21.66 $\pm$ 0.02 &   0.091 $\pm$ 0.002 &   0.348 $\pm$ 0.006 &   0.207 $\pm$ 0.003 &   0.795 $\pm$ 0.014 &  0.147 $\pm$ 0.006 &   0.655 $\pm$ 0.023 &   8.00 $\pm$ 0.30 &   0.74 $\pm$ 0.01 & 0.613 & 0.103 & -0.111\\
1442-0209\tnote{r} & 1237655498671849789 &   21.04 $\pm$ 0.01 &   -20.74 $\pm$ 0.01 &   0.055 $\pm$ 0.001 &   0.215 $\pm$ 0.002 &   0.123 $\pm$ 0.007 &   0.485 $\pm$ 0.003 &  0.093 $\pm$ 0.003 &   0.408 $\pm$ 0.011 &   8.00 $\pm$ 0.34 &   0.81 $\pm$ 0.02 & 0.388 & 0.392 & -0.099 \\
1454+4528 & 1237662301900964026 &   20.87 $\pm$ 0.02 &   -20.86 $\pm$ 0.02 &   0.058 $\pm$ 0.001 &   0.219 $\pm$ 0.005 &   0.150 $\pm$ 0.003 &   0.567 $\pm$ 0.012 &  0.117 $\pm$ 0.006 &   0.480 $\pm$ 0.024 &   8.00 $\pm$ 0.51 &   0.84 $\pm$ 0.03 & 0.691 & 0.303 & -0.057 \\
1457+2232 & 1237665549967294628 &   20.61 $\pm$ 0.01 &   -19.06 $\pm$ 0.01 &   0.075 $\pm$ 0.001 &   0.193 $\pm$ 0.002 &   0.163 $\pm$ 0.002 &   0.420 $\pm$ 0.004 &  0.240 $\pm$ 0.009 &   0.622 $\pm$ 0.024 &   8.00 $\pm$ 0.23 &   0.99 $\pm$ 0.01 & 0.237 & 0.347 & 0.159 \\
1503+3644\tnote{r} & 1237661872417407304 &   21.27 $\pm$ 0.01 &   -20.46 $\pm$ 0.01 &   0.056 $\pm$ 0.001 &   0.250 $\pm$ 0.003 &   0.109 $\pm$ 0.001 &   0.485 $\pm$  0.006 &  0.132 $\pm$ 0.006 &   0.660 $\pm$ 0.026 &   8.00 $\pm$ 0.33 &   0.80 $\pm$ 0.02 & 0.031 & 0.111 & -0.201 \\
1514+3852 & 1237661362380734819 &   21.02 $\pm$ 0.02 &   -20.52 $\pm$ 0.02 &   0.049 $\pm$ 0.001 &   0.191 $\pm$ 0.005 &   0.119 $\pm$ 0.003 &   0.466 $\pm$ 0.011 &  0.133 $\pm$ 0.009 &   0.637 $\pm$ 0.037 &   8.00 $\pm$ 0.50 &   0.67 $\pm$ 0.02 & 0 & 0.157 & -0.164 \\
1543+3446 & 1237662336790036887 &   21.21 $\pm$ 0.02 &   -18.72 $\pm$ 0.02 &   0.191 $\pm$ 0.004 &   0.475 $\pm$ 0.012 &   0.337 $\pm$ 0.007 &   0.839 $\pm$ 0.021 &  0.345 $\pm$ 0.039 &   1.081 $\pm$ 0.097 &   5.35 $\pm$ 0.33 &   0.63 $\pm$ 0.02 & 0 & 0.214 & 0.087 \\
1559+0841 & 1237662636912280219 &   21.69 $\pm$ 0.03 &   -19.63 $\pm$ 0.03 &   0.050 $\pm$ 0.002 &   0.209 $\pm$ 0.006 &   0.096 $\pm$ 0.003 &   0.397 $\pm$ 0.011 &  0.128 $\pm$ 0.016 &   0.568 $\pm$ 0.064 &   8.00 $\pm$ 0.95 &   0.88 $\pm$ 0.05 & 0 & 0.282 & -0.105 \\
2237+1336 & 1237656495641788638 &   20.18 $\pm$ 0.02 &   -21.76 $\pm$ 0.02 &   0.222 $\pm$ 0.003 &   0.966 $\pm$ 0.013 &   0.453 $\pm$ 0.006 &   1.966 $\pm$ 0.025 &  0.608 $\pm$ 0.058 &   2.669 $\pm$ 0.251 &   6.42 $\pm$ 0.35 &   0.98 $\pm$ 0.02 & 0.521 & 0.418 & -0.099 \\

\hline \hline \\
\label{tab1}

\end{tabular}
{\small
\begin{tablenotes}
\item[a] Green Pea IDs match those in Y17.
\item[b] SDSS DR14 BestObjID.
\item[c] The measured apparent UV magnitude.
\item[d] The associated errors are magnitude measurement uncertainties based on photon counting statistics (i.e., Poisson statistics) and propagation of the errors during the image calibration procedures such as flat-field correction.
\item[e] The extinction and $k$-corrected absolute UV magnitude at 1877 $\angstrom$.
\item[f] The circularized Petrosian half-light radius in units of arcseconds.
\item[g] The associated errors are based on the associated flux measurement uncertainties.
\item[h] The circularized Petrosian half-light radius in units of kpc assuming the adopted cosmological parameters in Section \ref{sec:introduction}.
\item[i] The circularized Petrosian radius in units of arcseconds.
\item[j] The circularized Petrosian radius in units of kpc assuming the adopted cosmological parameters in Section \ref{sec:introduction}.
\item[k] {The circularized effective radius and the associated uncertainties in units of arcseconds derived from the \textsf{GALFIT} surface brightness fitting in Section \ref{subsec:Size and Luminosity measurements}.}
\item[l] {The circularized effective radius in units of kpc assuming the adopted cosmological parameters in Section \ref{sec:introduction}.}
\item[m] {The \sersic\ index and the associated uncertainties derived from the \textsf{GALFIT} surface brightness fitting in Section \ref{subsec:Size and Luminosity measurements}.}
\item[n] The apparent axis ratio and the associated uncertainties derived from the \textsf{GALFIT} surface brightness fitting in Section \ref{subsec:Size and Luminosity measurements}.
\item[o] The internal extinction correction derived in {Paper I}
  based on the Balmer decrement. Zero indicates that the observed ratio of \Ha / \Hb is smaller than 2.86. See their section 2.3 and also Section 2.2 of {Y17} for details.
\item[p] The Milky Way extinction correction derived in {Paper I} based on the NASA/IPAC Galactic Dust Reddening and Extinction tool. See their section 2.3 for details.
\item[q] The $k$-correction derived in {Paper I} at 1877 $\angstrom$ assuming the UV slope of -2. See their section 2.3 for details.
\item[r] Confirmed Lyman-continuum leakers identified by \cite{izot16}.

\end{tablenotes}
}
\end{threeparttable}
\end{table*}

\begin{table*}[ht]
\centering
\begin{threeparttable}
\caption{The Representative UV Size {($r_{\rm{cir,eff}}$)} of GP Subsamples with Different $L_{\rm UV}$ and \ewlya\ Criteria}
\begin{tabular}{|l|rll|rll|}
\hline \hline
 $L_{\rm UV}/L_{\rm *,z=3}$ & N & Mean & Median & N & Mean & Median\\
 & \multicolumn{3}{l}{~~(No \ewlya\ Cut)} 
 & \multicolumn{3}{l}{~~(\ewlya\ $> 20 \ \angstrom$)}\\
\hline
{0.1}--0.3 &  10 &   0.68 &   0.60 $^{+0.19}_{-0.09}$ kpc\tnote{a} 
&  6 &  0.56 &  0.57 $^{+0.02}_{-0.11}$ kpc\tnote{a} \\
0.3--1 &  18 &   0.71 &   0.64 $^{+0.36}_{-0.16}$ kpc\tnote{a} 
& 16 &   0.67 &  0.64 $^{+0.27}_{-0.16}$ kpc\tnote{a} \\
1--{2.4} &  12 &  1.01 &  0.65 $^{+0.63}_{-0.10}$ kpc\tnote{a} &
  5 & 0.57 &  0.55 $^{+0.06}_{-0.01}$ kpc\tnote{a}\\
\hline
{0.1}--2.4 & 40 &  0.79 &  0.62 $^{+0.34}_{-0.15}$ kpc\tnote{a} 
& 27 &  0.63 &  0.57 $^{+0.13}_{-0.09}$ kpc\tnote{a} \\
\hline 
\end{tabular}
\label{tab2}
{\small
\begin{tablenotes}
\item[a] {The plus and minus values indicate the 75 th and 25 th percentiles from the median size, respectively.}
\end{tablenotes}
}
\end{threeparttable}
\end{table*}

\acknowledgments
We thank the referee for constructive comments that improved the manuscript. K.J.K. thanks Nathaniel R. Butler, Sanchayeeta Borthakur, and Rolf A. Jansen for helpful discussions. This work has been supported by HST-GO-14201 and HST-GO-15614 from STScI, which is operated by the Association of Universities for Research in Astronomy, Inc., for NASA under contract NAS 5-26555; and by NASA through Award No. NNG16PJ33C.


\clearpage
\end{document}